\documentclass[prd,twocolumn,floatfix,amsmath,nofootinbib,amssymb,floatfix]{revtex4}
\usepackage{graphicx,color,dcolumn,booktabs,bm}
\usepackage{longtable,lscape}
\usepackage{txfonts}
\usepackage{overpic}
\usepackage{amssymb}
\usepackage{indentfirst}
\usepackage{feynmf}   
\usepackage{slashed}  
\usepackage{cases}
\usepackage{color}
\usepackage{multirow}
\usepackage{epstopdf}
\usepackage{enumerate}
\usepackage{graphicx,color,dcolumn,booktabs,bm}
\usepackage[colorlinks, citecolor=blue,anchorcolor=red,menucolor=red, linkcolor=red,filecolor=red,urlcolor=blue,frenchlinks=red]{hyperref}

\graphicspath{{Figures/}} %

\begin{document}

\title{Fully-heavy structures in the invariant mass spectrum of $J/\psi \psi(3686)$, $J/\psi \psi(3770)$, $\psi(3686) \psi(3686)$, and $J/\psi \Upsilon(1S)$ at hadron colliders }
\author{Jun-Zhang Wang$^{1,2}$}\email{wangjzh2012@lzu.edu.cn}
\author{Xiang Liu$^{1,2,3}$\footnote{Corresponding author}}\email{xiangliu@lzu.edu.cn}
\author{Takayuki Matsuki$^{4}$}\email{matsuki@tokyo-kasei.ac.jp}
\affiliation{$^1$School of Physical Science and Technology, Lanzhou University, Lanzhou 730000, China\\
$^2$Research Center for Hadron and CSR Physics, Lanzhou University $\&$ Institute of Modern Physics of CAS, Lanzhou 730000, China\\
$^3$Lanzhou Center for Theoretical Physics, Lanzhou University, Lanzhou, Gansu 730000, China\\
$^4$Tokyo Kasei University, 1-18-1 Kaga, Itabashi, Tokyo 173-8602, Japan}

\date{\today}

\begin{abstract}
Motivated by a recent successful dynamical explanation for the newly observed fully-charm structure $X(6900)$ in the mass spectrum of di-$J/\psi$ by LHCb [J.~Z.~Wang \textit{et al.} arXiv:2008.07430], in this work, we extend the same dynamical rescattering mechanism to predict the line shape of more potential fully-heavy structures in the invariant mass spectrum of $J/\psi \psi(3686)$, $J/\psi \psi(3770)$, $\psi(3686) \psi(3686)$, and $J/\psi \Upsilon(1S)$ at high energy proton-proton collisions, whose verification in experiments should be helpful to further clarify the nature of $X(6900)$. The above final states of vector heavy quarkonia can be experimentally reconstructed more effectively by a $\mu^+\mu^-$ pair in the muon detector compared with $Q\bar{Q}$ meson with other quantum numbers.  Furthermore, the corresponding peak mass positions of each of predicted fully-heavy structures are also given. Our theoretical studies here could provide some valuable information for the future measurement proposals of LHCb and CMS, especially based on the accumulated data after completing Run III of LHC in the near future.
\end{abstract}
\maketitle

\section{\label{sec1}Introduction}

Since the first observation of charmonium $J/\psi$ in 1974 \cite{Aubert:1974js,Augustin:1974xw}, the fully-heavy-flavor physics has been one of the most hottest issues in the field of quantum chromodynamics(QCD). Benefitting from the unique heavy quark symmetry and non-relativistic behavior, the fully-heavy system is usually treated as an excellent platform to solve the non-perturbative puzzles of QCD \cite{Brambilla:2010cs,Brambilla:2004wf}. In the past several decades, within the continuous efforts from high energy collision experiments, a number of heavy quarkonium and quarkoniumlike states were discovered, especially novel charmoniumlike $XYZ$ structures, whose properties have provoked theoretical wide discussions and indeed largely enrich our knowledge for the color confinement interaction (see review articles in Refs. \cite{Chen:2016qju,Liu:2019zoy,Guo:2017jvc,Olsen:2017bmm,Brambilla:2019esw} for detail).

Although great progress has been made in the study of heavy quarkonium physics, both experimental and theoretical investigations for the fully-heavy systems containing beyond two heavy 
flavors are still absent.  Very recently, the LHCb Collaboration reported the measurements of double $J/\psi$ production by using proton-proton data at center of mass energies of 7, 8, and 13  TeV, where a clear peak around 6.9 GeV called 
$X(6900)$ and two underlying structures near the production threshold of $J/\psi J/\psi$ at 6.2 GeV and 7.3 GeV were observed in the invariant mass spectrum of di-$J/\psi$ \cite{Aaij:2020fnh}, respectively.
From the final states of $J/\psi J/\psi$, the observation of $X(6900)$ together with other two peaks indicates the first experimental evidence for fully-charm structures by the interaction of four charm flavors. Thus, the LHCb's observation has recently stimulated theorists with a great deal of enthusiasm to discuss the nature of $X(6900)$ in Refs. \cite{Chen:2020xwe,Jin:2020jfc,Lu:2020cns,Yang:2020rih,Deng:2020iqw,Wang:2020ols,Chen:2020lgj,liu:2020eha,Albuquerque:2020hio,Sonnenschein:2020nwn,Giron:2020wpx,
Richard:2020hdw,Becchi:2020uvq,Bedolla:2019zwg,Wang:2020wrp,Karliner:2020dta,Maciula:2020wri,Dong:2020nwy,Ma:2020kwb,Gordillo:2020sgc,Faustov:2020qfm,
Weng:2020jao,Zhang:2020xtb,Zhu:2020xni,Guo:2020pvt,Zhu:2020snb,Gong:2020bmg,Wan:2020fsk}, most of which contribute to the resonance interpretation of compact tetraquark hadronic state.

Among the recent theoretical explanations, different from the opinions of fully-charm tetraquark state, the Lanzhou group has proposed a dynamical mechanism to explore several new structures observed by LHCb \cite{Wang:2020wrp}. The key idea is based on a special dynamical contribution in reaction $pp \to J/\psi J/\psi X$, in which different combinations of a double charmonium directly produced in high energy proton-proton collisions are transferred into final state $J/\psi J/\psi$. Compared with a continuous distribution in the invariant mass spectrum of a $J/\psi$-pair, 
this mechanism has been found to produce an obvious cusp at the corresponding mass threshold of an intermediate double charmonium. By a theoretical analysis for the line shape of experimental data, three peaks observed by LHCb are well reproduced near 6.5, 6.9, and 7.3 GeV in the invariant mass spectrum of di-$J/\psi$, which naturally correspond to three rescattering channels of $\eta_c(1S) \chi_{c1}(1P)$, $\chi_{c0}(1P) \chi_{c1}(1P)$, and $\chi_{c0}(1P) X(3872)$ \cite{Wang:2020wrp}, respectively. From this point of view, the $X(6900)$ may not be a genuine resonance, which should be emphasized before declaring the discovery of a new exotic hadron state.

Of course, if the proposed dynamical explanation for $X(6900)$ \cite{Wang:2020wrp} is reasonable, we can naturally conjecture that this mechanism could be universal in the production of a double charmonium in high energy hadron colliders. Thus, the verification of this novel dynamical mechanism could be achieved by measurements of the different double charmonium production. Based on this motivation, in this work, we study possible fully-charm structures existing in high energy reactions, $pp \to J/\psi \psi(3686) X$, $pp \to J/\psi \psi(3770) X$, $pp \to \psi(3686) \psi(3686) X$ by the dynamical rescattering mechanism. In addition, we also predict a specific fully-heavy structures induced by the rescattering channels of a charmonium plus a bottomonium in the production process of $J/\psi \Upsilon(1S)$. Here, the reason for choosing vector charmonia $\psi(3686)$ and $\psi(3770)$ is that they belong to the same $J/\psi$ family and the feasibility of their prompt production has been proven in the high energy proton-proton collision experiments \cite{Aaij:2011sn,Aaij:2019evc}. Compared with other charmonium states with different quantum numbers, these charmonia together with bottomonium $\Upsilon(1S)$ can be experimentally reproduced more easily by the final states of $\mu^+ \mu^-$. Hence, we expect that the predictions presented in this work should be valuable to search for more fully-heavy structures in the invariant mass spectrum of a double charmonium and also could be tested in the future LHCb and CMS experiment.

This paper is organized as follows. After Introduction, we will present our theoretical framework how to calculate the dynamical rescattering contributions in the hadroproduction of a double heavy quarkonium in Sec.~\ref{sec2}. In Sec.~\ref{sec3}, the line shape predictions for potential fully-heavy structures on the invariant mass spectrum of $J/\psi \psi(3686)$, $J/\psi \psi(3770)$, $\psi(3686) \psi(3686)$, and $J/\psi \Upsilon(1S)$ are shown based on the dynamical rescattering mechanism, whose behaviors and peak positions are also discussed. This paper ends with the summary in Sec. \ref{sec4}.

\section{Dynamical rescattering mechanism in the hadroproduction of a double heavy quarkonium }\label{sec2}

\begin{figure}[b]
	\includegraphics[width=8.7cm,keepaspectratio]{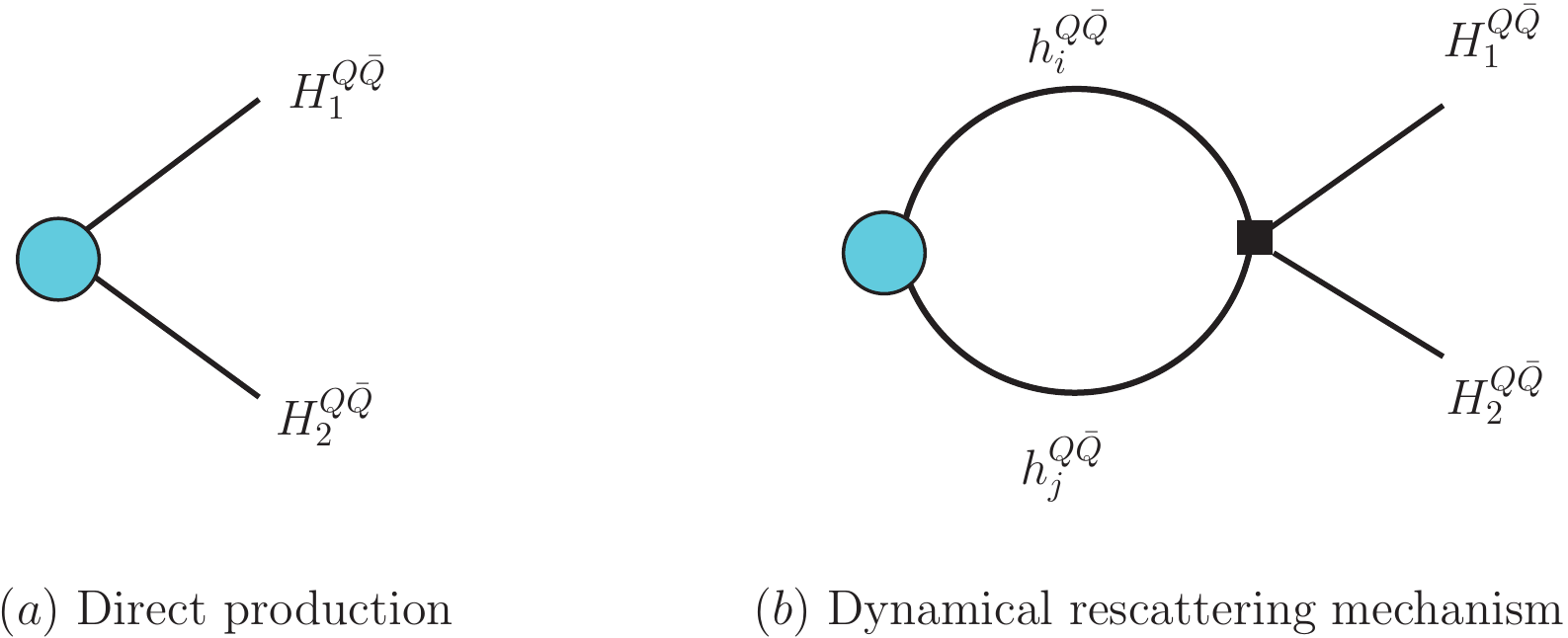}
	\caption{ The schematic diagrams for hadroproduction of a double heavy quarkonium marked by $H_1H_2$. Left diagram ($a$): the direct production process by single and double parton scattering; right diagram ($b$): the dynamical rescattering mechanism involving the various allowed intermediate heavy quarkonium pairs $h_ih_j$. The skybule circle represents direct production of a double heavy quarkonium in hadron collisions. \label{fig:schematic}  }
\end{figure}

The hadroproduction of a double heavy quarkonium in high energy proton-proton collisions is a very important subject of heavy quarkonium physics. At present, we know that a double heavy quarkonium can be directly produced by both the single parton scattering (SPS) and double parton scattering (DPS) processes \cite{Sun:2014gca,Likhoded:2016zmk,Baranov:2011zz,Lansberg:2013qka,Lansberg:2014swa,Lansberg:2015lva,Shao:2012iz,Shao:2015vga,Lansberg:2019adr,Calucci:1997ii,
Calucci:1999yz,DelFabbro:2000ds}. However, the situation may be more complicated in a real production process and some unknown dynamical effects may exist, where an available approach to test a possible underlying dynamical mechanism is the measurement of the corresponding invariant mass spectrum of a double heavy quarkonium. Focusing on a general production process $pp \to H_1H_2 X$, where $H_1H_2$ are the studied double heavy quarkonium. As shown in the schematic diagrams in Fig. \ref{fig:schematic}, in addition to the dominant direct production via SPS and DPS, the rescattering reaction of $ pp \to (h_ih_j \to H_1H_2) X$ may be an important underlying dynamical mechanism involved in the production of $H_1H_2$. Here, the intermediate particles $h_ih_j$ are composed of the combination of alternative double heavy quarkonium allowed by the system's quantum numbers of rescattering process $h_ih_j \to H_1H_2$. It is worth emphasizing that because of the lack of experimental information, our present knowledge for inner interaction of process $h_ih_j \to H_1H_2$ is still limited, so the coupling among intermediate charmonium pairs $h_ih_j$ and $H_1H_2$ has to be absorbed into a vertex for the convenience of the subsequent theoretical treatment.

Starting from an $S$-wave interaction between an intermediate heavy quarkonium pair $h_ih_j$, the production amplitude of $H_1H_2$ by dynamical rescattering mechanism becomes the one proportional to the scalar two-point loop integral, whose expression can be given by, in the rest frame of $H_1H_2$,
\begin{eqnarray}
&&L_{ij}(m_{H_1H_2})=\int \frac{dq^4}{(2\pi)^4} \frac{e^{-(2\vec{q}~)^{2}/\alpha^2}}{(q^2-m_{i}^2+i\epsilon)((P-q)^2-m_{j}^2+i\epsilon)} \nonumber \\
&&~=\frac{i}{4m_{i}m_{j}}
\left\{\frac{-\mu\alpha}{\sqrt{2}(2\pi)^{3/2}}+\frac{\mu\sqrt{2\mu m_0}\left(\textrm{erfi}\left[\frac{\sqrt{8\mu m_0}}{\alpha}\right]-i\right)}{2\pi/e^{-\frac{8\mu m_0}{\alpha^2}}}
\right\}, \label{eq:1}
\end{eqnarray}
where $\mu=(m_im_j)/(m_i+m_j)$ and $m_0=m_{H_1H_2}-m_{i}-m_{j}$. Here, $m_i$ and $m_j$ are the resonant mass of intermediate charmonium states $h_i$ and $h_j$, respectively, and $m_{H_1H_2}^2=(p_{H_1}+p_{H_2})^2$ is the square of the invariant mass of $H_1H_2$. $P=(m_{H_1H_2},0,0,0)$ represents the four-momentum of a $H_1H_2$ system and the erfi the imaginary error function. We also introduce an exponential form factor $e^{-(2\vec{q}~)^{2}/\alpha^2}$ to avoid the ultraviolet divergence of scalar two-point loop integral, and $\alpha$ is a cutoff parameter.

\begin{figure}[b]
	\includegraphics[width=8.7cm,keepaspectratio]{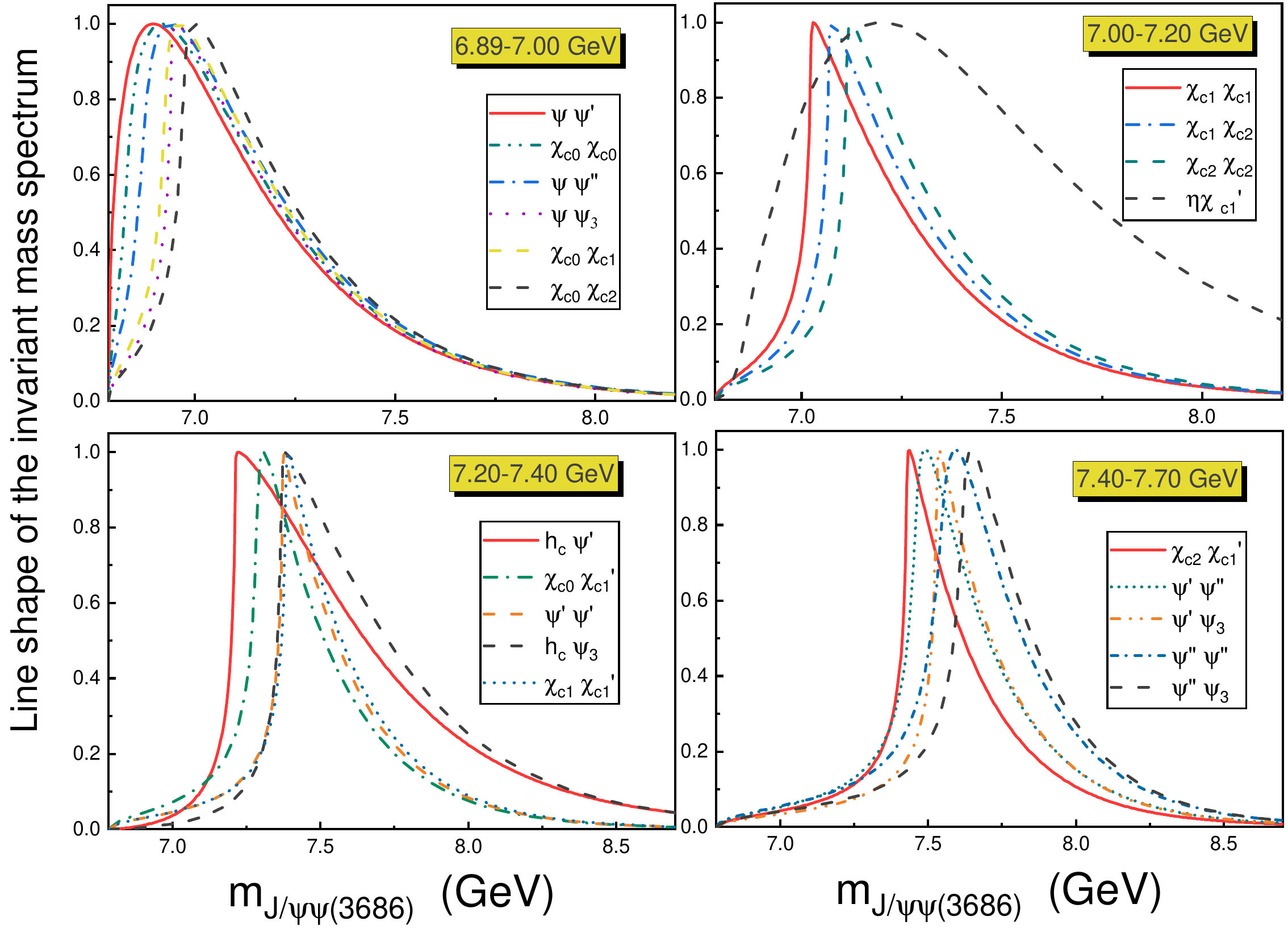}
	\caption{ The predicted line shapes of the invariant mass distribution of $J/\psi \psi(3686)$ produced in high energy proton-proton collisions using the only contributions from dynamical rescattering mechanism.  \label{fig:psiprimejpsi}  }
\end{figure}

In this work, we mainly consider the quantum number combination of $J^{PC}J^{PC}=1^{--}$$1^{--}$ for $H_1H_2$. Then, we can easily conclude that the system of $h_ih_j$ must satisfy $C=+1$ due to the conservation of $C$ parity. Based on this restriction, we can select the following allowed channels, $1^{--}$$1^{--}$, $0^{++}$$0^{++}$, $0^{++}$$1^{++}$, $0^{++}$$2^{++}$, $1^{++}$$1^{++}$, $1^{++}$$2^{++}$, and $2^{++}$$2^{++}$, etc. for the $h_ih_j$ with parity $P=+1$, and $1^{--}$$1^{+-}$, $0^{-+}$$0^{++}$, $0^{-+}$$1^{++}$, $0^{-+}$$2^{++}$, etc. for the $h_ih_j$ with parity $P=-1$. For the rescattering processes with two kinds of $P$ parity, the line shapes on the invariant mass spectrum of $m_{H_1H_2}$ can be given by \cite{Wang:2020wrp}
\begin{eqnarray}
\mathcal{A}^2_{ij}(m_{H_1H_2})=g_{ij}^2 L_{ij}^2(m_{H_1H_2})\frac{e^{c_0m_{H_1H_2}}\sqrt{\lambda(m_{H_1H_2}^2,m_{H_1}^2,m_{H_2}^2)}}{2m_{H_1H_2}^2} \label{eq:2}
\end{eqnarray}
and
\begin{eqnarray}
\mathcal{A}^{\prime 2}_{ij}(m_{H_1H_2})=g_{ij}^{\prime 2} L_{ij}^{ 2}(m_{H_1H_2})\frac{e^{c_0^{\prime}m_{H_1H_2}}\lambda(m_{H_1H_2}^2,m_{H_1}^2,m_{H_2}^2)^{\frac{3}{2}}}{8m_{H_1H_2}^4}, \label{eq:3}
\end{eqnarray}
respectively, where $\lambda(x,y,z)=x^2+y^2+z^2-2xy-2xz-2yz$ is the K$\ddot{\textrm{a}}$llen function. The exponential form factors $e^{c_0^{(\prime)}m_{H_1H_2}}$ are introduced when we parameterize a direct production amplitude of an intermediate double heavy quarkonium, which refers to the treatment of experimental analysis of LHCb \cite{Aaij:2020fnh} because of the complexity and difficulty in the present theoretical calculations \cite{He:2019qqr,He:2015qya,Lansberg:2020rft,Lansberg:2019fgm,Li:2009ug}. Since there are no relevant experimental data to determine the magnitude of coupling constants $g_{ij}^{(\prime)}$, we adjust the values of $g_{ij}^{(\prime)}$ to normalize the maximum of the line shape of $\mathcal{A}^{(\prime)2}_{ij}(m_{H_1H_2})$ to be one.

For the one-loop rescattering processes formulated by Eqs. (\ref{eq:2}-\ref{eq:3}), there exists a square root branch point in scalar two-point integral $L_{ij}(m_{H_1H_2})$, $\sqrt{m_{H_1H_2}-m_{i}-m_{j}}$, where an integral singularity at the threshold of $m_{i}+m_{j}$ appears at the on-shell of two intermediate heavy quarkonium states. The threshold singularity causes a cusp exactly at the corresponding threshold  in the invariant mass distribution of $m_{H_1H_2}$. However, in an actual process, the sharpness of a threshold cusp may be weakened by the resonant width of intermediate heavy quarkonium states. This width effect may be important for describing the line shape of the invariant mass spectrum for $m_{H_1H_2}$ and may even change the peak position from the threshold. So, the width effect will be considered in the following calculations by replacing $m_{i}$ and $m_{j}$ in Eq. (\ref{eq:1}) with $(m_{i}-i\Gamma_{i}/2)$ and $(m_{j}-i\Gamma_{j}/2)$, respectively. Within the above preparations, we can directly predict the line shapes of threshold cusps from dynamical rescattering mechanism in the hadroproduction of $J/\psi \psi(3686)$, $J/\psi \psi(3770)$, $\psi(3686) \psi(3686)$, and $J/\psi \Upsilon(1S)$ and the corresponding peak positions can also be given. In the following, we will discuss them carefully.

\section{Numerical results and discussions}\label{sec3}

\subsection{Fully-charm structures on the invariant mass spectrum of $J/\psi \psi(3686)$, $J/\psi \psi(3770)$, and $\psi(3686) \psi(3686)$}

\begin{figure}[t]
	\includegraphics[width=8.5cm,keepaspectratio]{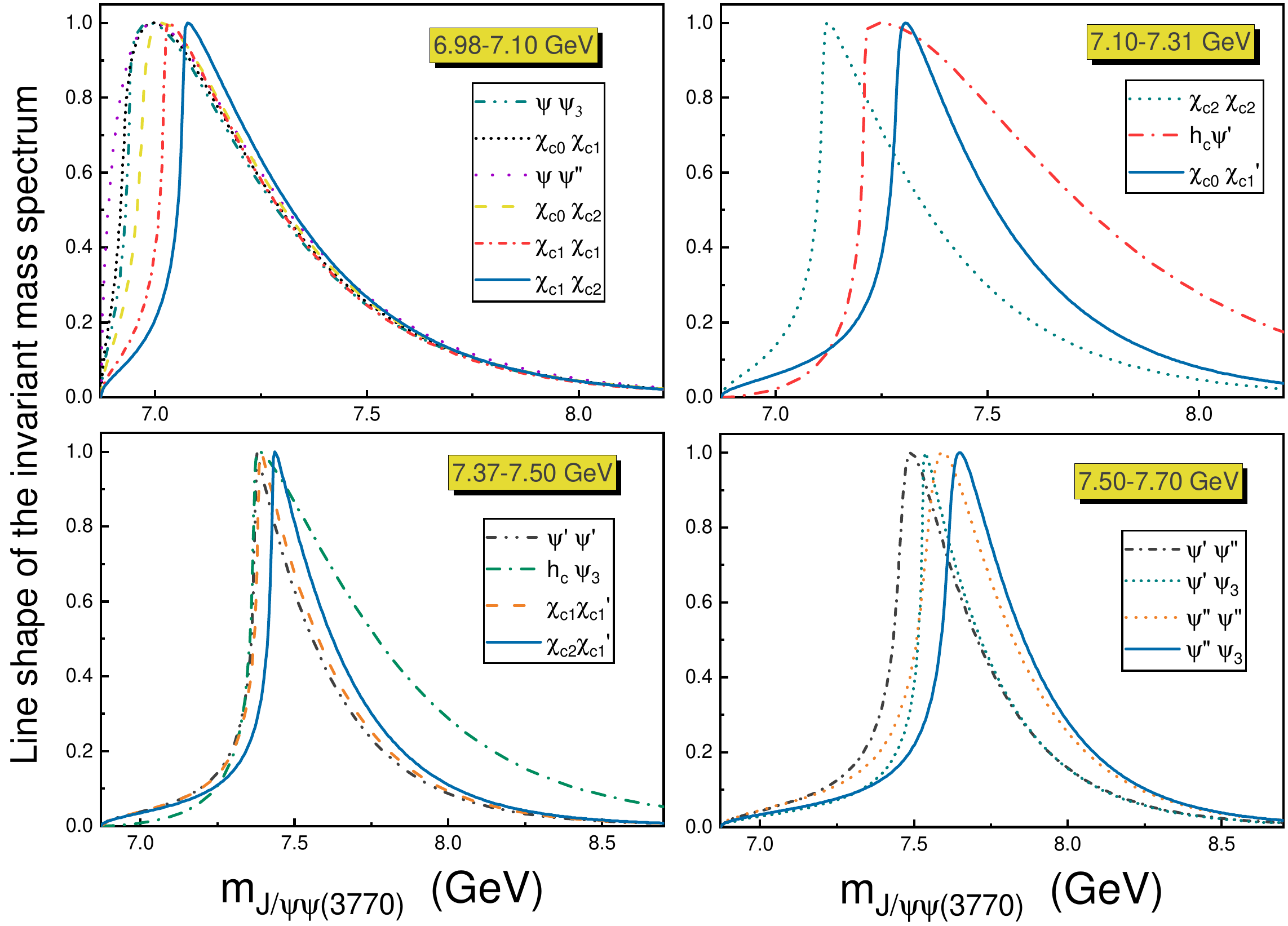}
	\caption{ The predicted line shapes of the invariant mass distribution of $J/\psi \psi(3770)$ produced in high energy proton-proton collisions using the only contributions from dynamical rescattering mechanism.  \label{fig:psi3770jpsi}  }
\end{figure}

\begin{figure}[t]
	\includegraphics[width=7.5cm,keepaspectratio]{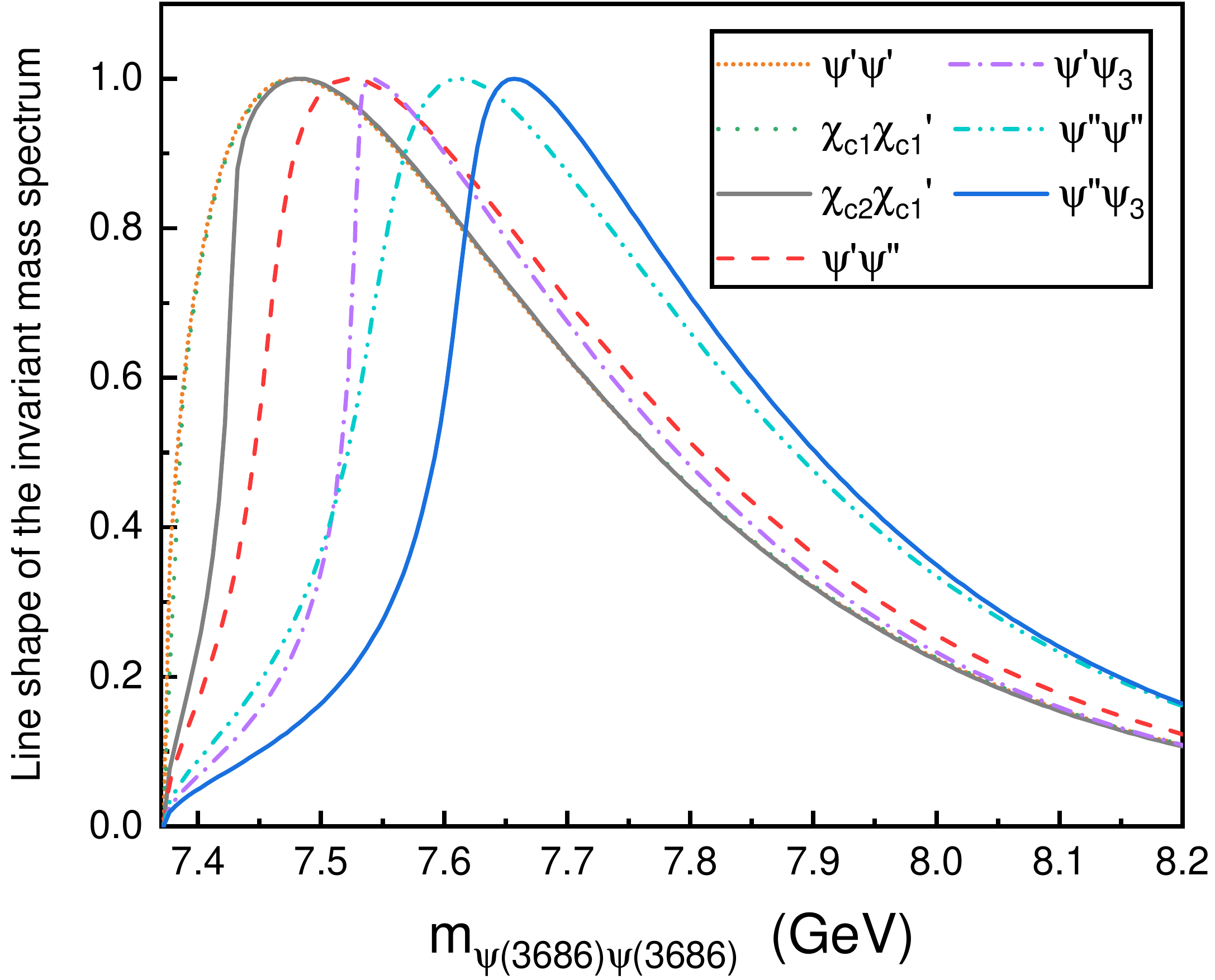}
	\caption{ The predicted line shapes of the invariant mass distribution of $\psi(3686) \psi(3686)$ produced in high energy proton-proton collisions using the only contributions from dynamical rescattering mechanism.  \label{fig:psiprimepsiprime}  }
\end{figure}

After a successful non-resonant dynamical explanation on $X(6900)$ observed in the invariant mass spectrum of $J/\psi J/\psi$ \cite{Wang:2020wrp}, we will extend our theoretical framework to discuss potential fully-charm structures in the hadroproduction of a double charmonium, $J/\psi \psi(3686)$, $J/\psi \psi(3770)$ and $\psi(3686) \psi(3686)$. According to the present charmonium spectroscopy \cite{Tanabashi:2018oca,Wang:2019mhs}, we select ten established charmonium or charmoniumlike states as intermediate rescattering particles in the dynamical mechanism, which are $\eta_c(1S)$($0^{-+}$), $J/\psi$, $\psi(3686)$, $\psi(3770)$($1^{--}$), $h_c(1P)$($1^{+-}$), $\chi_{c0}(1P)$, $\chi_{c1}(1P)$, $\chi_{c0}(1P)$, $X(3872)$, $\chi_{c2}(1P)$($J^{++}$ with $J=0,1,2$), and $X(3842)$($3^{--}$). Most of them have been directly discovered in high energy proton-proton experiments \cite{Aaij:2011sn,Aaij:2019evc,Aaij:2014bga,Aaij:2017tzn}. Additionally, it is worth emphasizing that the direct hadroproduction rates of $\eta_c$, $X(3872)$, and $P$-wave charmonium states $\chi_{cJ}$ with $J=0,1,2$ have been proven to be comparable with that of $J/\psi$ by both experiments \cite{Aaij:2014bga,Aaij:2011sn} and theoretical estimations from nonrelativistic QCD (NRQCD) \cite{Bodwin:1994jh,Ma:2014mri,Li:2011yc,Butenschoen:2014dra,Han:2014jya,Bodwin:2015iua,Ma:2010vd,Artoisenet:2009wk,Butenschoen:2013pxa}.
In the following discussions, without any special emphasis, $\eta_c$, $\psi$, $\psi^{\prime}$, $\psi^{\prime \prime}$, $h_c$, $\chi_{cJ}$, $\chi_{c1}^{\prime}$, and $\psi_3$ refer to $\eta_c(1S)$, $J/\psi$, $\psi(3686)$, $\psi(3770)$, $h_c(1P)$, $\chi_{cJ}(1P)$, $\chi_{c1}^{\prime}(2P)$ and $\psi_3(1D)$, respectively, and $\chi_{c1}^{\prime}(2P)=X(3872)$ \cite{Kalashnikova:2005ui,Zhang:2009bv,Kalashnikova:2009gt,Li:2009ad,Coito:2010if}, $\psi_3(1D)=X(3842)$ \cite{Wang:2019mhs}.

In this work, three model parameters $\alpha$, $c_0$, and $c_0^{\prime}$ are uniformly taken as 2.0, -1.5, and -1.0, respectively, which refers to the fitting results of the scenario-I for di-$J/\psi$ mass spectrum in Ref. \cite{Wang:2020wrp}. The predicted line shapes of fully-charm structures of the invariant mass distribution for $J/\psi \psi(3686)$ from high energy proton-proton collisions are presented in Fig. \ref{fig:psiprimejpsi}. It can be seen that there exist twenty allowed threshold cusps at the energy region from 6783 to 7700 MeV. These peak structures can be divided into four energy regions, i.e., ($6.89\sim7.00$), ($7.00\sim7.20$), ($7.20\sim7.40$) and ($7.40\sim7.70$) GeV as seen in Fig. \ref{fig:psiprimejpsi}. Furthermore, we find that the peaks of up to five channels $\psi \psi^{\prime}$, $\chi_{c0} \chi_{c0}$, $\psi \psi^{\prime \prime}$, $\psi \psi_3$, and $\chi_{c0} \chi_{c1}$ are clustered in a short energy region of 6900 to 6960 MeV, which are close to the production threshold of 6783 MeV. Simultaneously, there is also a similar energy region with small range of 7370 to 7440 MeV, which include four channels of $\psi^{\prime} \psi^{\prime}$, $h_c \psi_3$, $\chi_{c1} \chi_{c1}^{\prime}$ and $\chi_{c2} \chi_{c1}^{\prime}$.   At present experimental statistics, it is not likely to identify the individual signal from these close peaks and their contributions may overlap because of the similar line shapes near peaks. According to the above analyses, we strongly encourage experimentalists to search for the most promising two fully-charm structures on the invariant mass spectrum of $J/\psi \psi(3686)$ near 6.9 and 7.4 GeV. As for the remaining rescattering channels, it is  worth noting that the cusp effect from $\eta_c \chi_{c1}^{\prime}$ channel is relatively weak so its contribution may be covered by the direct production background.

The predicted line shapes of fully-charm structures on the invariant mass distributions of $J/\psi \psi(3770)$ and $\psi(3686) \psi(3686)$ from high energy proton-proton collisions are shown in Figs. \ref{fig:psi3770jpsi} and \ref{fig:psiprimepsiprime}, respectively. Here, the allowed intermediate double charmonium channels are consistent with those in $J/\psi \psi(3686)$. Because of a general suppression of the contributions from off-shell channels, we consider only the intermediate channels above a production threshold and then there are seventeen and seven selected rescattering channels for hadroproduction of $J/\psi \psi(3770)$ and $\psi(3686) \psi(3686)$, respectively. Influenced by the phase space distribution, their peak line shapes will be fatter than those in the invariant mass spectrum of $J/\psi \psi(3686)$, especially for the channels near the production threshold. Similarly to Fig. \ref{fig:psiprimejpsi}, we can see from Fig. \ref{fig:psi3770jpsi} that there also exist two typical energy regions of  $(6990 \sim 7040)$ MeV and $(7370 \sim 7440)$ MeV for the hadroproduction of $J/\psi \psi(3770)$, which are related to five double charmonium channels of $\psi \psi^{\prime\prime}$, $\psi \psi_3$, $\chi_{c0} \chi_{c1}$, $\chi_{c0} \chi_{c2}$, and $\chi_{c1} \chi_{c1}$ and four channels of $\psi^{\prime} \psi^{\prime}$, $h_c \psi_3$, $\chi_{c1} \chi_{c1}^{\prime}$, and $\chi_{c2} \chi_{c1}^{\prime}$, respectively. This means that it is worth expecting to observe two clear structures near 7.0 and 7.4 GeV in the invariant mass spectrum for $J/\psi \psi(3770)$ in the future LHCb and CMS experiments. As for the hadroproduction of $\psi(3686) \psi(3686)$, considering that its production threshold reaches 7372 MeV, we suggest the experiments to explore possible fully-charm structures near 7.5 GeV, which corresponds to five close threshold cusps of double charmonium channels for $\psi^{\prime} \psi^{\prime}$, $\chi_{c1} \chi_{c1}^{\prime}$, $\chi_{c2} \chi_{c1}^{\prime}$, $\psi^{\prime} \psi^{\prime\prime}$, and $\psi^{\prime} \psi_3$ as shown in Fig. \ref{fig:psiprimepsiprime}.

\begin{table}[t]
  	\caption{The peak mass positions of different rescattering channels in the invariant mass spectrum for $J/\psi \psi(3686)$, $J/\psi \psi(3770)$, and $\psi(3686) \psi(3686)$ in high energy proton-proton collisions. The results are all in unit of MeV.}
  	\setlength{\tabcolsep}{2.0mm}{
  	\begin{tabular}{cccccccc}
			\toprule[1.0pt]
            \toprule[1.0pt]
    Rescattering channels &  $m_{J/\psi \psi(3686)}$ & $m_{J/\psi \psi(3770)}$  & $m_{\psi(3686) \psi(3686)}$ \\
			\toprule[1.0pt]
          $\psi \psi^{\prime}$ & 6895  &  $\cdots$   &    $\cdots$       \\
          $\chi_{c0} \chi_{c0}$ & 6916  &  $\cdots$   &   $\cdots$         \\
          $\psi \psi^{\prime \prime}$ & 6939  & 6998    & $\cdots$          \\
          $\psi \psi_3$ & 6953  &  6987   &    $\cdots$       \\
          $\chi_{c0} \chi_{c1}$ & 6958  &  6996   &     $\cdots$       \\
          $\chi_{c0} \chi_{c2}$ &  6999 &  7016   &    $\cdots$        \\
          $\chi_{c1} \chi_{c1}$ &  7029 &  7032   &    $\cdots$       \\
          $\chi_{c1} \chi_{c2}$ &  7076 &  7078   &     $\cdots$      \\
          $\chi_{c2} \chi_{c2}$ &  7122 &  7124   &    $\cdots$       \\
          $\eta_c \chi_{c1}^{\prime}$ & 7198  &  $\cdots$   &    $\cdots$       \\
          $h_c \psi^{\prime}$ & 7222  &  7247   &    $\cdots$      \\
          $\chi_{c0} \chi_{c1}^{\prime}$ & 7306  &  7307   &    $\cdots$        \\
          $\psi^{\prime} \psi^{\prime}$ &  7375 &   7375  &    7478        \\
          $h_c \psi_3$ & 7383  &  7386   &   $\cdots$        \\
          $\chi_{c1} \chi_{c1}^{\prime}$ & 7389  &  7389   &    7479      \\
          $\chi_{c2} \chi_{c1}^{\prime}$ & 7436  &  7436   &    7482      \\
          $\psi^{\prime} \psi^{\prime \prime}$ & 7490  &  7491   &   7524     \\
          $\psi^{\prime} \psi_3$ & 7536  &  7536   &  7542      \\
          $\psi^{\prime \prime} \psi^{\prime \prime}$ & 7593  &   7593  &   7612    \\
          $\psi^{\prime \prime} \psi_3$ & 7648  & 7648    &   7657   \\
			\bottomrule[1.0pt]
		\end{tabular}\label{table1}}
  \end{table}

In addition to the line shapes of threshold cusps from different rescattering channels, the corresponding invariant mass positions at peaks 
are also given and summarized in Table \ref{table1}. For the intermediate channels composed of point-like particles, their peak mass positions should exactly equal to the mass summation of intermediate states. However, due to the width effects of intermediate resonances and the phase space distribution function, the actual peak mass position is generally larger than the threshold position. In addition, we find that though the line shape of a threshold cusp is obviously dependent on the model parameters $\alpha$ and $c_0^{(\prime)}$, they have little effects on the maximum position. Therefore, the predictions for the peak mass positions of different fully-charm structures listed in Table \ref{table1} should be credible as long as the masses and widths of intermediate charmonia are determined, which will be valuable for the future experimental search proposals.

\subsection{Fully-heavy structures involved with b-flavor in the invariant mass spectrum of $J/\psi \Upsilon(1S)$}

We can continue to extend the dynamical rescattering mechanism to the predictions of fully-heavy structures involved with bottom flavor. The potential fully-bottom structures in the hadroproduction of a double bottomonium $\Upsilon \Upsilon$ have been studied in Ref. \cite{Wang:2020wrp}. In this subsection, we will focus on a special case of fully-heavy structures, which will be hopefully discovered in the invariant mass distribution of a charmonium $J/\psi$ plus a bottomonium $\Upsilon(1S)$ from high energy proton-proton collisions. For the candidates of intermediate charmonia and bottomonia in the dynamical production processes of $J/\psi \Upsilon(1S)$, we consider only low-lying $\eta_c$, $\psi$, $h_c$, $\chi_{cJ}$ and $\eta_b$, $\Upsilon$, $h_b$, $\chi_{bJ}$ with $J=0,1,2$, respectively, where unknown widths of several bottomonium states are taken by theoretical estimations \cite{Wang:2018rjg}.  Here, it is worth emphasizing that though the present studies on bottomonium production in the high energy proton-proton experiments are still absent, but the observation of high excited states $\chi_{b1}(3P)$ and $\chi_{b2}(3P)$ by CMS in 2018 \cite{Sirunyan:2018dff} has proven the ability of LHC to produce the $b\bar{b}$ states. Thus, the measurements of the hadronic production of bottomonia are still worth being expected in the future.

\begin{figure}[t]
	\includegraphics[width=8.7cm,keepaspectratio]{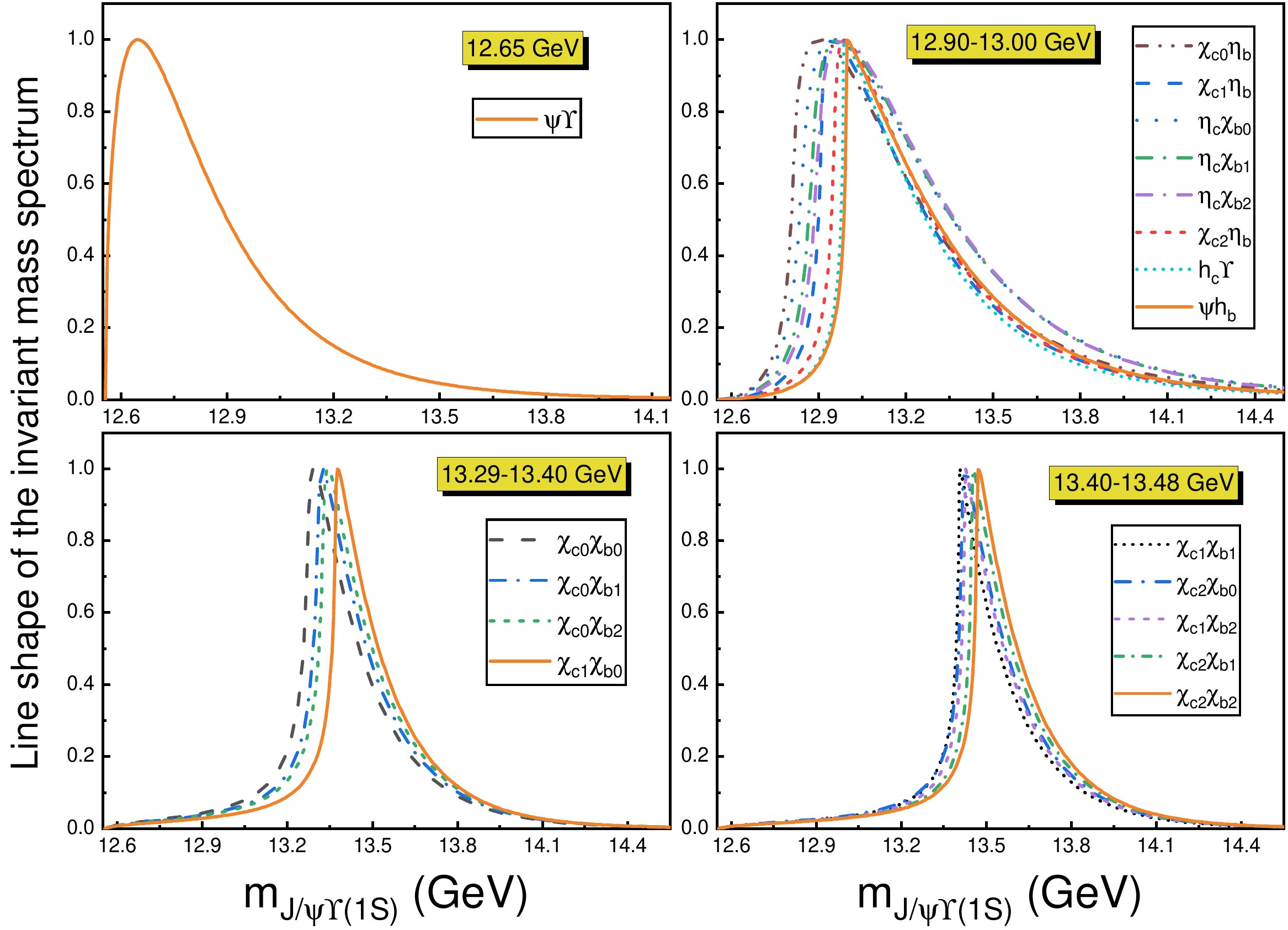}
	\caption{ The predicted line shapes of the invariant mass distribution of $J/\psi \Upsilon(1S)$ produced in high energy proton-proton collision using the only contributions from dynamical rescattering mechanism.  \label{fig:upsilonjpsi}  }
\end{figure}

\begin{table}[t]
  	\caption{The peak mass positions of different rescattering channels in the invariant mass spectrum of $J/\psi \Upsilon(1S)$ in high energy proton-proton collision.}
  	\setlength{\tabcolsep}{10.0mm}{
  	\begin{tabular}{cccccccc}
			\toprule[1.0pt]
            \toprule[1.0pt]
    Rescattering channels &  $m_{J/\psi \Upsilon(1S)}$ (MeV) \\
			\toprule[1.0pt]
          $\psi \Upsilon$ & 12648         \\
          $ \chi_{c0} \eta_b$ & 12903          \\
          $ \chi_{c1} \eta_b$ & 12948          \\
          $\eta_c \chi_{b0}$ & 12952        \\
          $\eta_c \chi_{b1}$ & 12968       \\
          $\eta_c \chi_{b2}$ &  12981      \\
          $\chi_{c2} \eta_b$ &  12990      \\
          $h_c \Upsilon$ &  12990      \\
          $\psi h_b$ &  12998      \\
          $\chi_{c0} \chi_{b0}$ & 13293        \\
          $\chi_{c0} \chi_{b1}$ & 13324       \\
          $\chi_{c0} \chi_{b2}$ & 13343          \\
          $\chi_{c1} \chi_{b0}$ & 13378         \\
          $\chi_{c1} \chi_{b1}$ & 13407         \\
          $\chi_{c2} \chi_{b0}$ & 13425      \\
          $\chi_{c1} \chi_{b2}$ & 13426       \\
          $\chi_{c2} \chi_{b1}$ & 13454    \\
          $\chi_{c2} \chi_{b2}$ & 13473      \\
			\bottomrule[1.0pt]
		\end{tabular}\label{table2}}
  \end{table}

The predicted line shapes and peak mass positions of fully-heavy structures in the invariant mass distribution of $J/\psi \Upsilon(1S)$ from high energy proton-proton collisions are shown in Fig. \ref{fig:upsilonjpsi} and Table \ref{table2}, respectively. Benefited from heavy quark symmetry, which causes an approximate mass degeneracy between two $S$-wave bottomonia $\eta_b$ and $\Upsilon$ as well as among four $P$-wave bottomonium states $\chi_{bJ}$ with $J=0,1,2$ and $h_b$, the threshold cusps in the invariant mass spectrum of $J/\psi \Upsilon(1S)$ are mainly concentrated in four separate energy regions, i.e., $12.65$, ($12.90\sim13.0$), ($13.29\sim13.40$) and ($13.40\sim13.48$) GeV as shown in Fig. \ref{fig:upsilonjpsi}. Here, it can be seen that a near-threshold structure at 12.65 GeV is provided by the channel of $\psi \Upsilon$. Anyway, the above four energy regions are highly recommended for future relevant experimental measurements.

\section{Summary}\label{sec4}

Recently, the LHCb collaboration reported the observation of a new structure $X(6900)$ in the reconstruction events of di-$J/\psi$, which is the first evidence for the existence of fully-heavy structures in the invariant mass distributions of a double heavy quarkonium \cite{Aaij:2020fnh}. On account of the importance of $X(6900)$ discovery, its nature has aroused great interests among theorists. In Ref. \cite{Wang:2020wrp}, we have proposed a special dynamical mechanism to explain the peak line shape of $X(6900)$, whose core is a dynamical rescattering process that the allowed combinations of an intermediate double charmonium directly produced in high energy proton-proton collisions are transferred into a final state $J/\psi J/\psi$. Furthermore, we have found that these processes could produce the obvious threshold cusps near the position of mass summation of the corresponding intermediate double charmonium.

Motivated by a successful description of experimental line shapes for di-$J/\psi$ mass spectrum of LHCb by a dynamical rescattering mechanism \cite{Wang:2020wrp}, in this work, we have extended our theoretical framework to study more fully-charm structures in the invariant mass spectrum of a different double charmonium from high energy proton-proton collisions, which are $J/\psi \psi(3686)$, $J/\psi \psi(3770)$, and $\psi(3686) \psi(3686)$. According to our theoretical predictions, we have strongly recommended some hopefully detectable fully-charm structures in the invariant mass spectrum of a double charmonium to experimentalists, whose peak mass positions are 6.9 and 7.4 GeV for $J/\psi \psi(3686)$, 7.0 and 7.4 GeV for $J/\psi \psi(3770)$, and 7.5 GeV for $\psi(3686) \psi(3686)$, respectively. A special case of fully-heavy structures involved with bottom flavor in the hadroproduction of $J/\psi \Upsilon(1S)$ has been also predicted, which are found to be clustered in four energy regions of $12.65$, ($12.90\sim13.0$), ($13.29\sim13.40$), and ($13.40\sim13.48$) GeV.

Just like the observation of a fully-charm structure $X(6900)$ \cite{Aaij:2020fnh}, when the couplings between intermediate rescattering channels and products of a double charmonium are strong enough, it is easy to distinguish the peak signals of threshold cusps from the direct production background by SPS and DPS mechanisms, which usually behave like a continuous distribution. Fortunately, in near future, the Run III of LHC will be performed and then the High-Luminosity-LHC upgrade will achieve a data collection of an integrated luminosity of 300 fb$^{-1}$ in pp collisions at a CM energy of 14 TeV \cite{Bediaga:2018lhg}. Therefore, we greatly expect that these novel fully-heavy structures predicted in this work can be observed in the future measurements, especially at LHCb and CMS.

\section*{ACKNOWLEDGEMENTS}

This work is partly supported by the China National Funds for Distinguished Young Scientists under Grant No. 11825503, the National Program for Support of Top-notch Young Professionals and the 111 Project under Grant No. B20063.


\end{document}